\documentclass[preprint,aps,nofootinbib]{revtex4}
\usepackage{framed}
\usepackage{bbold}
\usepackage{slashed}
\usepackage{graphicx}
\usepackage{amsmath}
\usepackage{amssymb}
\usepackage{epsfig}
\usepackage{hhline}
\usepackage{soul}
\usepackage{tabularx,pifont,multirow}
\usepackage{enumitem}
\usepackage{colordvi}
\usepackage{soul}
\usepackage{xcolor}
\usepackage{yfonts}

\newcommand{\be}{\begin{equation}}
\newcommand{\ee}{\end{equation}}

\begin{document}
\begin{titlepage}

\begin{flushleft}
{\small KCL-PH-TH/2019-{\bf 42}}
\end{flushleft}

\vspace{0.1cm}

\begin{center}
{\it \Large \bf Do we Come from a Quantum Anomaly? \footnote{This essay received an {\em Honorable Mention} from the Gravity Research Foundation Essay
Competition on Gravitation (2019)}}

\vskip 0.3cm { \bf Spyros Basilakos$^{a,b}$, Nick E. Mavromatos$^{c}$ and Joan Sol\`a Peracaula$^d$}
\end{center}

\begin{quote}
\begin{center}
$^a$Academy of Athens, Research Center for Astronomy and Applied Mathematics, Soranou Efessiou 4, 115 27 Athens, Greece. \\
$^b$ National Observatory of Athens, Lofos Nymfon,
11852, Athens, Greece. \\
$^c$Theoretical Particle Physics and Cosmology Group, Physics Department, King's College London, Strand, London WC2R 2LS, UK. \\
 $^{d}$Departament de F\'\i sica Qu\`antica i Astrof\'\i sica, and Institute of Cosmos Sciences (ICCUB), Univ. de Barcelona, Av. Diagonal 647 E-08028 Barcelona, Catalonia, Spain.
\end{center}
\end{quote}

\vspace{0.2cm}
{\small E-mails: svasil@academyofathens.gr, nikolaos.mavromatos@kcl.ac.uk, sola@fqa.ub.edu}

\vspace{0.2cm} \centerline{\small (Submission date: March 28, 2019)} \vspace{0.5cm}
\centerline{\bf Abstract}
\bigskip
We present a string-based picture of the cosmological evolution in which  (CP-violating) {\it gravitational anomalies} acting during the inflationary phase of the universe cause the vacuum energy density to ``run'' with the effective Hubble parameter squared, $H^2$,  thanks to the axion field of the bosonic string multiplet. This leads to baryogenesis through leptogenesis with massive right-handed neutrinos. The generation of chiral matter after inflation helps in cancelling the anomalies in the observable radiation- and matter- dominated eras. The present era inherits the same ``{\it running vacuum}'' structure  triggered  during the inflationary time by the axion field.  The current dark energy is thus predicted to be mildly dynamical, and dark matter should be made of axions.   Paraphrasing  Carl Sagan\,\cite{sagan}: {\it we are all anomalously made from starstuff} .


\vspace{0.2cm}


\end{titlepage}

\pagestyle{plain} \baselineskip 0.75cm

\section{Gravity and chiral anomalies in an anomaly-free Universe}

``$A\nu\omega\mu\alpha\lambda\iota\alpha$'' (in English: ``Anomaly'') is a word of Greek origin, used to describe the ``deviation from the standard, normal or expected situation''. In (Quantum) Physics, anomalies are associated with the breaking of symmetries at quantum level, in the sense that the corresponding currents, which are classically conserved, no longer are so in the quantum theory. The most common one is the chiral anomaly\, \cite{AdlerBellJackiw}.  Gauge theories also suffer from anomalies,  leading to the loss of gauge covariance at a quantum level unless they cancel\,\cite{GaumeGinspark1985}.

Anomalies, far from being ``errors'' or  ``arbitrariness''  emerging from the laws of Nature, are a  consequence of them.  They are  implied by the principles inherent to the  Quantum Theory of the World. The first anomalies in the Universe were the gravitational anomalies and they were responsible for inflation, but after that they remained unbalanced  (``out of order'').  The material Universe, however, abhors to be ``anomalous''.  That  is why (quantum) matter came to the existence, only to provide a second generation of anomalies (the chiral ones) able to neutralize the {first}.  Thanks to quantum matter the Friedman-Lema\^\i tre-Robertson-Walker (FLRW) universe is anomaly free (hence  ``in order''!) and classical General Relativity (GR) can describe it. Such is the ultimate ``raison d'\^etre'' of matter (and of ourselves!) --  or at least this is the main thesis of this essay.

Gravitational anomalies are the origin of this story \cite{grav,jackiw,GaumeGinspark1985}. They can  appear e.g. as one-loop Feynman diagrams with external graviton legs, in which chiral fermions circulate in the loop.  At first sight  they appear as a sickness of the theory.  We are used to Einstein's GR, in which Physics does not depend on the coordinates of space-time, a property referred to as general covariance.  Gravitational anomalies can break such symmetry.  But this is not a tragedy,  if matter is not present. Matter appearance can actually restore general covariance  at the quantum level and the  standard -- anomaly-free -- FLRW cosmology may exist.

In this Essay, we support this idea through a concrete, albeit simplified  model  inspired from string theory, in which the early inflationary phase  possesses only the massless degrees of freedom of the (bosonic) gravitational string multiplet: a {massless} spin-2 tensor  field, which is uniquely identified as the graviton, a scalar (spin-0) field, the dilaton, and a spin-one antisymmetric tensor, $B_{\mu\nu}=-B_{\nu\mu}$, whose field strength (in four space-time dimensions) is dual  to a pseudoscalar massless  field $b(x)$ (Kalb-Ramond (KR) axion). The latter actually plays a key  r\^ole in this story, as the presence of the anomaly and its coupling to KR axions leads to  an (approximately) constant kinetic energy for $b(x)$  over the inflationary period. This in turn leads to  contributions to the vacuum energy density,  which adopt the  {\it running-vacuum} form  $\rho_{\Lambda}\sim H^2$ ~\cite{rvm1,rvm2,rvm3,bls}, with $H$ the (approximately) constant Hubble rate during inflation, with a coefficient which is determined by the slow-roll parameter $\epsilon$.
During the subsequent radiation era, the non-diluted  presence of the axion field $b(x)$  enables  leptogenesis to occur through the CP-violating decays of, say, massive right handed neutrinos to conventional particles~\cite{cms,bms}. The so generated lepton asymmetry can then be communicated to baryons~\cite{strumia} via the usual sphaleron interactions in the SM sector~\cite{krs}, thus explaining the baryon asymmetry in the universe.

In the current Universe, gravitational anomalies are creeping in anew, as in any evolving stage approaching de Sitter.  The KR axion can have a similar form as in the inflationary era and triggers once more  $H^2$-contributions to $\rho_{\Lambda}$.  The gravitational anomalies become eventually exposed, but  again this is harmless (as it was in the Universe's infancy) since matter is fading away.  In the interlude, the current DM is axion-like and the  DE is not just a rigid $\Lambda=$const.  but a constant plus  $\sim H^2$ with a small coefficient: $\rho_\Lambda=c_0+\nu\,H^2$.  The DE is thus predicted dynamical, which  is {favoured} by current observations and can alleviate the $\Lambda$CDM tensions\,\cite{EPL2018,MNRAS2018,GoSolBas2015}.
Everything that follows is a summarized attempt to justify this thesis\,\cite{Fullpaper}.


\section{Primordial Gravitational Waves Induce Anomalies}

The essential bosonic part of the (four-space-time-dimensional) effective action, $S_B$, that reproduces the string scattering amplitudes to lowest non trivial order reads~\cite{gsw,string}
\be\label{sea}
S_B  =\; \int d^{4}x\sqrt{-g}\Big( \dfrac{1}{2\kappa^{2}} [-R + 2\, \partial_{\mu}\Phi\, \partial^{\mu}\Phi] - \frac{1}{6\kappa^2}\, e^{-4\Phi}\, { H}_{\lambda\mu\nu}{H}^{\lambda\mu\nu}
+ \dots \Big),
\ee
where $\dots$ represent higher derivative terms ({we ignore for our purposes here vacuum energy terms that may appear in various forms in the context of string/brane models. Their presence does not affect our main arguments.}). Here  ${ H}_{\nu\rho\sigma}$ is  the field strength of the antisymmetric tensor field  $B_{\mu\nu}$, which satisfies the Bianchi identity $\partial_{[\mu}\, {H}_{\nu\rho\sigma]} = 0$ (a  generalization of the electromagnetic case);  $\kappa=1/M_{\rm Pl}$
with $M_{\rm Pl}=M_P/\sqrt{8\pi} = 2.43 \times 10^{18}$~GeV the reduced Planck mass.  In this work, we shall assume that the dilaton varies slowly. This implies an (approximately) constant string coupling $g_s = g_s^{(0)} e^{\Phi_0}=g_s^{(0)}$ upon setting $\Phi_0 = 0$.

In the presence of gauge and gravitational fields, cancelation of anomalies requires the modification of the field strength ${ H}_{\mu\nu\rho}$ by appropriate gauge (Yang-Mills) and Lorentz Chern--Simons terms~\cite{gsw}. As a {result,} the modified (anomalous) Bianchi identity mentioned above becomes
\begin{equation}\label{modbianchi2}
 \varepsilon_{abc}^{\;\;\;\;\;\mu}\, \nabla_\mu \, {H}^{abc}=  \frac{\alpha^\prime}{32} \, \sqrt{-g}\, \Big(R_{\mu\nu\rho\sigma}\, \widetilde R^{\mu\nu\rho\sigma} -
F_{\mu\nu}\, \widetilde F^{\mu\nu}\Big) \equiv \sqrt{-g}\, {\mathcal G}(\omega, \mathbf{A})\,,
\end{equation}
where $\nabla_\mu$ denotes gravitational covariant derivative.The tildes denote the dual tensors, as usual. The anomaly ${\mathcal G}(\omega, \mathbf{A})$  can be expressed as a total derivative, $\partial_\mu \Big(\sqrt{-g} \, {\mathcal K}^\mu (\omega) \Big)$, in which ${\mathcal K}^\mu (\omega)$ depends on the spin-connection $\omega^a_{\,\,b}$.  Integrating over the ${ H}$ field in the path integral of the bosonic string action \eqref{sea}, one obtains the effective action\,\cite{kaloper}
\begin{align}\label{sea4}
S^{\rm eff}_B =\int d^{4}x\sqrt{-g}\Big[ -\dfrac{1}{2\kappa^{2}}\, R + \frac{1}{2}\, \partial_\mu b(x) \, \partial^\mu b(x)  -
 \sqrt{\frac{2}{3}}\,
\frac{\alpha^\prime}{96 \, \kappa} \, \partial_\mu b(x) \, {\mathcal K}^\mu + \dots \Big],
\end{align}
{where} $\dots$ {denote} higher derivative terms, and {for our purposes in this work we set}\footnote{{This is a choice, given that the string scale is in general different from the four-dimensional Planck mass scale. We consider the simplest scenario here, where any compactification radius of the underlying string theory is close to Planck scale.}}:
$\alpha^\prime \sim \kappa^2$.
The axion field $b(x)$ {will} be a main actor in this story, as it couples to the gravitational and gauge fields via  CP- violating interactions.

While gravitation anomalies are absent in the FLRW background, they appear when (primordial) gravitational-waves exist during the inflationary phase.
Taking into account the CP-violating derivative coupling of the KR axion to the anomaly current ${\mathcal K}^\mu$ in \eqref{sea4},
one finds, up to second order in graviton fluctuations and to leading order in $k \, \eta \gg 1$ ~\cite{stephon}:
\begin{align}\label{rrt}
  \langle R_{\mu\nu\rho\sigma}\, \widetilde R^{\mu\nu\rho\sigma} \rangle  = \frac{16}{a^4} \, \kappa^2\, \int \frac{d^3k}{(2\pi)^3} \, \frac{H^2 }{2\,k^3} \, k^4 \, \Theta + {\rm O}(\Theta^3) ,
 \end{align}
where $\eta$ ({respectively} $t$) is the conformal (cosmic) time in the de Sitter era: $\eta=H^{-1} \exp(-Ht)$, with $H=$const.,
 $k$ being the Fourier mode,  $a$ the scale factor; and finally
 $\Theta = \sqrt{\frac{2}{3}}\, \frac{\alpha^\prime \, \kappa}{12} \, H \,  {\dot {b}} \, \ll \, 1$, {with the
 dot} denoting $d/dt$. The integral \eqref{rrt} is restricted to physical modes $k/a$ satisfying  $k\, \eta  < \mu /H,$ where  $\mu$ is a UV  cutoff.

 The crucial point now is that  \eqref{rrt} violates  general covariance. As noted, this should not be viewed as a `catastrophe' in the absence of matter (see next section).
From (\ref{sea4}), the classical field equation of  $b(x)$ obtains:
\begin{align}\label{krbeom2}
\partial_{\alpha}\Big[\sqrt{-g}\Big(\partial^{\alpha}\bar{b}  -  \sqrt{\frac{2}{3}}\,
\frac{\alpha^\prime}{96 \, \kappa} \, {\mathcal K}^{\alpha}  \Big)\Big] = 0 \, \Rightarrow \, \boxed{\dot{\overline{b}}  =  \sqrt{\frac{2}{3}}\, \frac{\alpha^\prime}{96 \, \kappa} \, {\mathcal K}^{0}\sim {\rm const}}\,.
\end{align}
thus implying the existence of a background $\overline b (t)$ that respects the (large scale) homogeneity and isotropy,
with a slow-roll evolution for the anomaly:
\begin{eqnarray}\label{k02}
{\mathcal K}^0 (t)
 \simeq {\mathcal K}^0_{\rm begin} (0) \, \exp\Big[  - 3H\, t \, \Big( 1  -  0.73 \,  \times 10^{-4} \,  \Big(\frac{H}{M_{\rm Pl}}\Big)^2 \, \Big(\frac{\mu}{M_{\rm Pl}}\Big)^4 \Big)\Big].
\end{eqnarray}
The slow-roll is a consequence of the  CMB measurements~\cite{planck}, which during inflation imply  $\frac{H}{M_{\rm Pl}}\,  \in \, \Big[ 10^{-5} , 10^{-4} \Big)$.  However,  for
$\frac{H}{M_{\rm Pl}} = \Big(10.8062 \, \frac{M_{\rm Pl}}{\mu}\Big)^2$, the  exponent in \eqref{k02} {\it vanishes} and the anomaly current is $\sim${\it constant} during inflation, hence {\it undiluted} at its exit.  For  phenomenologically acceptable ranges of $H \ll M_{\rm Pl}$~\cite{planck},  {\it transplankian modes} ($\mu\gg  M_{\rm Pl}$) should  thus be involved.


To estimate  ${\mathcal K}_{\rm begin}^0(t=0)$, we assume that the rates of the KR-axion and of the inflaton $\varphi$  are of the same order of magnitude, the only constraint being ${\dot {\overline b}}\ll{\mathcal U}(\varphi)|$  so as not to upset the slow-roll condition:  $\epsilon = \frac{1}{2} \frac{1}{(H M_{\rm Pl})^2}\, {\dot \varphi}^2 \sim \frac{1}{2} \frac{1}{(H M_{\rm Pl})^2}\, {\dot {\overline b}}^2 \sim 10^{-2}$\,~\cite{planck}.
Whence
 \begin{align}\label{slowrollepsi}
 \boxed{{\dot {\overline b}} \sim  \sqrt{2\,\epsilon} \, M_{\rm Pl} \, H \sim  0.14 \, M_{\rm Pl} \, H}~ .
 \end{align}
Using \eqref{krbeom2}, the (approximately constant, during inflation) anomaly ${\mathcal K}^0 \sim {\mathcal K}_{\rm begin} (t=0)$  reads  ${\mathcal K}^0 \sim {\mathcal K}_{\rm begin} (t=0) \sim 16.6 \, H \, M_{\rm Pl}^2$;
and with the help of the $b$-field stress tensor and  \eqref{slowrollepsi}  the anomaly input to the energy density appears in the fashion of {\it running vacuum} corrections~\cite{rvm1,rvm2}:
 \begin{align}\label{enpressphib2}
 \rho^{\varphi+b} \simeq 3M_{\rm Pl}^4 \Big[3.33  \times 10^{-3} \, \Big(\frac{H_{\rm infl}}{M_{\rm Pl}}\Big)^2  + \frac{{\mathcal U}(\varphi)}{3M_{\rm Pl}^4}\Big].
  \end{align}
Inflation occurs as long as ${\mathcal U}(\varphi) \gg 10^{-2} \, (H_{\rm infl}\, M_{\rm Pl})^2 $.

\section{Cancelling gravitational anomalies in the matter era}

After the exit from inflation, gravitational anomalies could jeopardize the incipient radiation phase.
But the KR-axion comes to rescue!  {First, it allows to erase the offending gravitational part, and  second it
provides the anomalous chiral coupling  to chiral fermionic matter, essential for leptogenesis~\cite{cms,bms}.  Indeed, the new modified effective action (\ref{sea4}) reads}
\begin{align}\label{sea6}
S^{\rm eff} =&\; \int d^{4}x\sqrt{-g}\Big[ -\dfrac{1}{2\kappa^{2}}\, R + \frac{1}{2}\, \partial_\mu b \, \partial^\mu b +  \kappa\, \,  b(x)  \, \nabla_\mu \, \Big(\sqrt{\frac{2}{3}}\,
\frac{1}{96} \, {\mathcal K}^\mu - \frac{1}{2} \, \sqrt{\frac{3}{2}} \,  J^{5 \, \mu}\Big)
\, \Big] + \dots,
\end{align}
{and involves the chiral current $J^{5 \, \mu} = \sum_{j} \, \bar{\psi}_j\,  \gamma^{\mu} \,\gamma^{5}\psi_j$. Its {gravitational covariant four-divergence} may just {\it cancel}  the offending gravitational  anomaly ${\mathcal K}^\mu$ in \eqref{sea6} and  restore  general covariance at quantum level,  leaving only the mentioned chiral anomaly part (harmless for general covariance, of course):}
\begin{align}\label{cancel}
\nabla_\mu \, \Big(\sqrt{\frac{3}{8}} \,  {J}^{5\, \mu} - \sqrt{\frac{2}{3}}\, \frac{1}{96} \, {\mathcal K}^\mu \Big) = { ``{chiral~anomalies}''}.
\end{align}
The  post inflationary equation of motion for the KR axion,
$ \partial_\mu \Big(\sqrt{-g} \, \partial^\mu b(x) \Big) =  {``{chiral~anomalies''}}$,   leaves  a bulk axion  flow  ($\dot{\overline b}  \propto T^3 + { subleading~terms}$, {slowly varying with temperature $T$ during leptogenesis}~\cite{bms})  which defines our very source of  dark matter.

{We explain briefly at this stage how the appearance of the undiluted  Lorentz- and CPT-({\it spontaneously})-violating KR background \eqref{slowrollepsi} at the end of the inflationary period leads to phenomenologically consistent leptogenesis~\cite{Fullpaper,bms}  in models with {\it massive} right-handed sterile neutrinos.
Indeed, the presence of such a background,  implies that the latter
affects the decays of the massive sterile neutrinos into standard model particles, through Higgs-portal interactions linking the standard-model leptonic sector with the sterile-neutrino
one. This is a consequence of the non-trivial coupling of the KR axion with the four divergence of the axial current, which for the sterile neutrinos
is classically not conserved as a result of their non-trivial mass. This leads to the generation of a leptonic asymmetry during the radiation era, at a freeze-out temperature of order of the mass of the (lightest of the sterile) neutrinos~\cite{Fullpaper}. In the scenario of \cite{bms}, this mass is of order of $10^5-10^7$~GeV, thus compatible with Higgs mass stability during the electroweak symmetry breaking~\cite{Vissani}. We do not discuss here how the sterile neutrino mass is generated. One way, in which gravitational anomalies play a role,
is described in \cite{pilaftsis}, and involves mixing of the KR axion with other axions, that are abundant in string models, which interact with the sterile neutrinos via chirality changing Yukawa couplings.}

Back to the main traits of the cosmological evolution in our framework, in the current era we re-encounter  an (approximate) de Sitter  phase,  carrying gravitational anomalies due to gravity-wave perturbations. The KR-axion may {have~\cite{Fullpaper}} a similar slow-roll regime as in the inflationary era \eqref{slowrollepsi}, with running  $H^2$-contributions to the  vacuum energy (\ref{enpressphib2}). However, being the current $H_0\ll H_{\rm inf}$, the gravitational anomalies are rendered as innocuous as in the early Universe.  Furthermore, as in (\ref{slowrollepsi}) we expect
 \begin{equation}\label{bdot0}
 \boxed{{\dot b}_{\rm today} \sim  \sqrt{2\,\epsilon^\prime} \, M_{\rm Pl} \, H_0}~.
 \end{equation}
With a  potential $U_b$ giving mass to $b(x)$, it  is quite tempting to assume $\epsilon^\prime \sim \epsilon$  and identify the DM with the axion background left after inflation.
Note from \eqref{bdot0} and \eqref{slowrollepsi},
that the slow-roll parameter of $b(x)$ measures the ratio of its kinetic energy, $K_b\sim (1/2)\,\dot{b}^2$,  to the critical energy density of the Universe, $\rho_c=(M_{\rm Pl} H)^2/3$. Thus,  on identifying $\epsilon^\prime=\epsilon \sim 10^{-2}$ and estimating {that} $K_b$ is roughly one order of magnitude smaller than $U_b$ now ({a typical situation of other cosmological fields, such as e.g. quintessence), we can get the dark matter content in the right ballpark ~\cite{planck}:
\begin{equation}\label{eq:UbTb}
  \Omega_m=\frac{\rho_m}{\rho_c}\simeq \frac{U_b}{\rho_c}\simeq 10\frac{K_b}{\rho_c}\simeq 10\epsilon={\cal O}(0.1)\,.
\end{equation}

In short:  the (gravitational) anomaly played an important dual r\^ole for  {\it our existence}: first, it induced a non-diluted axion background of DM at the end of inflation into the radiation epoch, {which itself induces leptogenesis}; and,
second,  it fostered the subsequent generation of chiral  matter  from the decay of the running vacuum, thus cancelling the unbalanced gravitational anomaly and restoring GR in our Universe; leaving also a (mildly) running DE\,\cite{EPL2018,MNRAS2018,GoSolBas2015} -- as a smoking gun {of it}!

So, indeed,  `\emph{we might well be anomalously made of starstuff!}

\vspace{-0.5cm}

\section*{Acknowledgements}

\vspace{-0.5cm} 

SB acknowledges support from
the Research Center for Astronomy of the Academy of Athens in the
context of the program  ``{\it Tracing the Cosmic Acceleration}''.The work of NEM is supported in part by STFC (UK) under the research grant ST/P000258/1.The work of JSP has
been partially supported by projects  FPA2016-76005-C2-1-P (MINECO), 2017-SGR-929 (Generalitat de Catalunya) and MDM-2014-0369 (ICCUB). 
This work is also partially supported by the COST Association Action CA18108 ``{\it Quantum Gravity Phenomenology in the Multimessenger Approach (QG-MM)}''.
NEM acknowledges a scientific associateship (``\emph{Doctor Vinculado}'') at IFIC-CSIC-Valencia University, Valencia, Spain.

\end{document}